\documentstyle[12pt,aasms4]{article}

\lefthead{Umemoto et al.}
\righthead{Ammonia in L1157}

\begin{document}

\title{The ortho-to-para ratio of ammonia in the L1157 outflow \altaffilmark{1}}

\author{Tomofumi UMEMOTO \altaffilmark{2}, Hitomi MIKAMI \altaffilmark{3}, Satoshi YAMAMOTO\altaffilmark{4}, \\and Naomi HIRANO\altaffilmark{5}}

\altaffiltext{1}{Based on observations made at the Nobeyama Radio Observatory (NRO). 
Nobeyama Radio Observatory is a branch of the National Astronomical Observatory, 
an inter-university research institute operated by the Ministry of Education, Science, Sport, and  Culture, Japan.}

\altaffiltext{2}{National Astronomical Observatory, Mitaka, Tokyo, 181-8588, JAPAN}

\altaffiltext{3}{Nobeyama Radio Observatory, Minamimaki, Minamisaku, Nagano, 384-1305, JAPAN}

\altaffiltext{4}{Department of Physics and Research Center for the Early Universe, 
The University of Tokyo, Bunkyo-ku, Tokyo, 113-0033, JAPAN}

\altaffiltext{5}{Laboratory of Astronomy and Geophysics, 
Hitotsubashi University,
Kunitachi, Tokyo, 186-8601, JAPAN}

\begin{abstract}
We have measured the ortho-to-para ratio of ammonia in the blueshifted
 gas of the L1157 outflow by observing the six metastable inversion lines 
from ($J, K$) = (1, 1) to (6, 6).
The highly excited (5, 5) and (6, 6) lines were first detected in the low-
mass star forming regions.
The rotational temperature derived from the ratio of four transition
lines from (3, 3) to (6, 6) is 130--140 K, suggesting that the 
blueshifted gas is heated by a factor of $\sim$10 as compared
to the quiescent gas.
The ortho-to-para ratio of the NH$_3$ molecules in the blueshifted gas is estimated to be 1.3--1.7, 
which is higher than the statistical equilibrium value.
This ratio provides us with evidence that the NH$_3$ molecules have been
 evaporated from dust grains 
with the formation temperature between 18 and 25 K.
It is most likely that the NH$_3$ molecules on dust grains have been 
released into the gas phase 
through the passage of strong shock waves produced by the outflow.
Such a scenario is supported by the fact that the ammonia abundance in 
the blueshifted gas is enhanced by a factor of $\sim$5 
with respect to the dense quiescent gas.
\end{abstract}


\section{INTRODUCTION}

The inversion lines of metastable ammonia have been extensively used 
for observations of dense ($\sim$10$^4$ cm$^{-3}$) molecular cloud 
cores (e.g. \markcite{Mye83,Ben89}Myers \& Benson 1983; Benson \& 
Myers 1989), 
while the fractional abundance of NH$_3$ varies among the clouds 
(\markcite{Ben83}Benson \& Myers 1983).
On the basis of a systematic survey, 
\markcite{Suz92}Suzuki et al. (1992) pointed out that the NH$_3$ tends 
to be more abundant in the star-forming cores than the starless cores.
Such a trend is interpreted in terms of the gas-phase chemical evolution 
which predicts that the NH$_3$ is deficient in the early stages of 
chemical evolution and becomes abundant in the later stages 
(e.g. \markcite{Her89,Suz92}Herbst \& Leung 1989; Suzuki et al. 1992).
In addition to the gas-phase chemical evolution, the gas-grain interaction 
is also considered to be important for the abundance variation of NH$_3$ 
(e.g. \markcite{dHe85,Bro90,Nej90,Flo95}d'Hendecourt, Allamandola, \& 
Greenberg 1985; Brown \& Charnley 1990; Nejad, Williams, 
\& Charnley 1990; Flower, Pineau des For\^{e}ts \& Walmsley 1995).
In the star-forming dense cores, the NH$_3$ molecules retained on dust 
grains can be released into gas phase by means of shock waves 
caused by the outflow and the radiation from the newly formed star (e.g. 
\markcite{Nej90,Flo95}Nejad et al. 1990; Flower et al. 1995), 
and are expected to contribute to increase the NH$_3$ abundance.

Usually, it is not easy to estimate contribution of molecules desorbed 
from dust grains from those formed in the gas phase.
However in the case of NH$_3$, an ortho-to-para ratio of NH$_3$ 
provides us with such information.
Ammonia has two distinct species, the ortho-NH$_3$ ($K = 3n$) and the 
para-NH$_3$ ($K \neq 3n$), which arise from different relative 
orientations of the three hydrogen spins.
The ortho-to-para ratio is expected to be the statistical equilibrium 
value of 1.0 
when the NH$_3$ molecules are formed in the processes of gas-phase 
reactions or grain surface reactions.
This is because the reactions to form the NH$_3$ molecule release large 
excess energies compared to the energy difference between the lowest 
ortho and para states, 
which make the ortho-to-para ratio close to the statistical value.
However, the ortho-to-para ratio is expected to be larger than unity when 
the NH$_3$ molecules adsorbed on grain surface released into the gas 
phase with the excess energy for desorption 
which is comparable to the energy difference between the ortho and para 
states.
This is because the lowest energy level of the para species is 23 K higher 
than that of the ortho species, 
para species require more energy for desorption than ortho species.
The time scale of the interconversion between ortho-NH$_3$ and para-
NH$_3$ is considered to be the order of 10$^6$ yr in the gas phase
(\markcite{Che69}Cheung et al. 1969).
Therefore, the ortho-to-para ratio provides us with valuable information 
on physical conditions and chemical processes 
when the NH$_3$ molecules are released into the gas phase.

Since the lowest energy level of ortho-NH$_3$, ($J, K$) = (0, 0), has no 
inversion doubling, 
it is necessary to observe the transitions higher than ($J, K$) = (3, 3) to 
measure the ortho-to-para ratio.
Such high transitions are hardly excited in the dark clouds of low 
temperature, however, recent observations have revealed that 
shocked gas associated with molecular outflows in the dark clouds is 
heated enough to excite the transitions higher than ($J, K$) = (3, 3).
One of the prototypical objects with shock-heated gas is the bipolar 
outflow in the dark cloud L1157 (e.g. \markcite{Ume92}Umemoto et al. 
1992) which is located at 440 pc from the sun (\markcite{Vio69}Viotti 1969).
Previous NH$_3$ observations have revealed that the gas in the outflow 
is heated to more than 50--100 K 
(\markcite{Bac93, Taf95}Bachiller, Mart\'{\i}n-Pintado, \& Fuente 1993; 
Tafalla \& Bachiller 1995).
A good morphological coincidence of the NH$_3$ distribution (\markcite
{Taf95}Tafalla \& Bachiller 1995) with that of the SiO ($J$=2--1) 
emission (\markcite{Zan95,Gue98}Zang et al. 1995; Gueth et al. 1998), 
which is considered to be a good tracer of the shocked molecular gas, 
reveals that the hot ammonia arises from the shocked gas.
The NH$_3$ abundance enhancement observed in the shocked gas 
(\markcite{Taf95}Tafalla \& Bachiller 1995) suggests the possibility 
that the NH$_3$ retained on grain mantles is released into the gas-phase 
by means of shocks.

In this {\it Letter}, we report the observations of six metastable 
inversion lines of NH$_3$ from ($J, K$) = (1, 1) to (6, 6) toward the 
blueshifted lobe of the L1157 outflow.
We have detected the high excitation NH$_3$ (5, 5) and (6, 6) lines for the 
first time in the low-mass star forming regions.
Of six observed transitions, the (3, 3) and (6, 6) states belong to ortho-
NH$_3$ and the other four states belong to para-NH$_3$.
The detection of both (6, 6) and (3, 3) emission enables us to measure the 
ortho-to-para ratio which provides us with information on the 
contribution of ammonia molecules desorbed from grains.

\section{OBSERVATIONS}

The observations were carried out in 1991 June and 1992 May with the 
45m telescope of the Nobeyama Radio Observatory.
We observed six inversion transitions of the metastabel ammonia, 
($J, K$) = (1,1), (2, 2), (3, 3), (4, 4), (5, 5), and (6, 6), 
toward three positions along the blueshifted lobe of the CO outflow. 
The position offsets from IRAS 20386+6751 (hereafter referred to as 
L1157 IRS), $\alpha$(1950) = 20$^{\rm h}$38$^{\rm m}$39.$^{\rm s}$6, 
$\delta$(1950) = 67$^{\circ}$51$'$33$''$, 
are (0$''$, 0$''$), (20$''$, -60$''$), and (45$''$, -105$''$), 
which were referred as positions A, B, and C, respectively by 
\markcite{Mik92}Mikami et al. (1992).
Positions B and C correspond to the tips of two CO cavities
 (\markcite{Gue96}Gueth, Guilloteau, \& Bachiller 1996),
toward which strong NH$_3$ (3, 3) and SiO ($J$=2--1) emission lines are 
observed 
(\markcite{Taf95,Zha95,Gue98}Tafalla \& Bachiller 1995; Zhang et al. 
1995; Gueth et al. 1998).
We also observed the (1, 1), (2, 2), and (3, 3) transitions toward 
additional two positions outside of the lobe 
and one position toward the peak of the redshifted CO emission.
The observed positions are shown in Figure 1 superposed on the 
CO ($J$=1--0) outflow map observed by 
\markcite{Ume92}Umemoto et al. (1992).
The lowest three transitions were observed simultaneously.
The higher three transitions were also observed simultaneously.
At the observed frequencies, the telescope had a beam size of 72$''$ and 
the beam efficiency of 0.8.
The frontend was a cooled HEMT receiver whose typical system noise 
temperature was $\sim$200 K.
The backend was a bank of eight acousto-optical spectrometers with 37 
kHz resolution, 
which corresponds to a velocity resolution of 0.50 km s$^{-1}$.
All the observations were made in position switching mode.
To obtain the spectra, we have integrated 45 minutes per position, 
the resulting rms noise level per channel was $\sim$ 0.02 K.

\section{RESULTS}

In Figure 2, we show the spectra of six transitions observed at positions 
A, B, and C.
At position A, a strong (1, 1) line and weak (2, 2) and (3, 3) lines are 
observed.
No emission from the transitions higher than (4, 4) has been detected.
The (1, 1) line observed at position A shows the quiescent component 
peaked at V$_{\rm LSR}$ = 2.85 km s$^{-1}$, 
which is almost the same as the cloud systemic velocity of 
$\sim$2.7 km s$^{-1}$ derived from the $^{13}$CO observations 
(\markcite{Ume92}Umemoto et al. 1992), with a weak blueshifted wing.
The (2, 2) and (3, 3) lines appear in the velocity range of the (1, 1) wing 
emission, 
suggesting that these line profiles are contaminated by the blueshifted 
component in the southern lobe because of the large beam size of 72$''$.
Since the (2, 2) and (3, 3) lines show no significant emission at the 
velocity of the quiescent component,
the rotational temperature, $T_{\rm rot}$, of the quiescent component is 
estimated to be $\lesssim$10 K.

At positions B and C, all six transitions up to (6, 6) were detected.
The (5, 5) and (6, 6) lines are first detected in the low-mass star forming regions.
All transition lines except the (1, 1) one have peak velocities blueshifted 
by 1--2 km s$^{-1}$ from the cloud systemic velocity 
and show broad widths of ${\Delta}V$ $\sim$4--10 km s$^{-1}$ 
(measured at 1$\sigma$ level), 
indicating that the emission of the higher energy level arises from the 
high velocity gas.
The (3, 3) line profiles at positions B and C resemble the profiles of the 
SiO (J=2-1) and CS (J=2-1) lines 
observed by \markcite{Mik92}Mikami et al. (1992).
Among the six transition lines, the (3, 3) line is the strongest as 
previously pointed out by \markcite{Bac93}Bachiller et al. (1993).
The peak brightness temperatures of the (2, 2), (3, 3), and (4, 4) lines 
observed at position B 
are lower by a factor of 1.5--2 than those observed by \markcite
{Bac93}Bachiller et al. (1993) at the same position.
This may be due to the beam dilution effect, because the spatial extent of 
the ammonia emitting area is smaller than our beam 
(\markcite{Bac93,Taf95} Bachiller et al. 1993; Tafalla \& Bachiller 1995).

Toward three other positions outside the blueshifted lobe, 
we have detected only the (1, 1) emission line at the systemic velocity.

\section{DISCUSSION}
\subsection{Temperature and Ortho-to-para Ratio of the High-Velocity Gas}

To obtain the rotational temperature $T_{\rm rot}$ of the blueshifted 
gas, we assumed optically thin emission, 
and constructed rotation diagrams, i.e., the logarithms of NH$_3$ column 
densities divided by statistical weight of the transition, 
plotted against the energy above the ground state.
The contribution of the quiescent component in the (1, 1) line was 
eliminated by performing the multi-components gaussian fitting.
The rotational temperature estimated from the (1, 1) and (2, 2) data, 
$T_{\rm rot}$(1,1; 2,2) is 38 K at positions B and is 48 K at position C. 
These temperatures are a factor of 4 to 5 higher than that of the 
quiescent component at position A.
Figure 3 shows that the $T_{\rm rot}$ obtained from the data of higher 
transitions are significantly higher than $T_{\rm rot}$(1,1; 2,2).
The rotational temperature obtained from the para-NH$_3$ (4, 4) and 
(5, 5) data, $T_{\rm rot}$(4,4; 5,5), are $\sim$130 K for both positions B and C.
The slopes between the ortho-NH$_3$ (3, 3) and (6, 6) data are almost 
parallel to those between the para-NH$_3$ (4, 4) and (5, 5) ones, 
indicating that $T_{\rm rot}$(3,3; 6,6) is comparable to $T_{\rm rot}$(4,4; 5,5).

It should be noted that the column densities of the ortho species are 
higher than those of para species; 
this suggests that the ortho species are more abundant than the para 
species.
If we assume an ortho-to-para ratio to be $\sim$1.5, the data of four 
transitions involving the ortho- and para-states align on the straight 
lines; 
the best fits provide an ortho-to-para ratio at position B to be 
1.7${{+0.2} \atop {-0.3}}$ and that at position C to be 1.3$\pm$0.2.
Then we obtain the $T_{\rm rot}$ at position B is 
140${{+4} \atop {-3}}$K (the upper and lower limit of the $T_{\rm rot}$ 
correspond to those of the ortho-to-para ratio, respectively) and that at 
position C is 125${{+3} \atop {-4}}$K.

The discrepancy between the $T_{\rm rot}$(1,1; 2,2) and the 
$T_{\rm rot}$ derived from the higher transition data 
can be explained by two components of gas with different kinetic 
temperatures as suggested by 
\markcite{Ave96}Avery \& Chiao (1996) from their SiO observations.
However, we consider that the most of the gas traced by the NH$_3$ 
emission is heated to 130--140 K in the following reason.
It is known that the metastable populations may deviate from a true 
Boltzmann distribution 
due to collisional depopulation of the higher nonmetastable levels.
As argued by \markcite{Dan88}Danby et al. (1988), 
the $T_{\rm rot}$ tends to underestimate the gas kinetic temperatures 
$T_{\rm k}$, at the $T_{\rm k}$ range higher than $\sim$ 30 K.
Since the difference is remarkable for the $T_{\rm rot}$ derived from the 
lower $J$ transitions, 
the $T_{\rm rot}$ estimated from the higher transitions is considered to 
be the better indicator of the kinetic temperature than 
$T_{\rm rot}$(1,1; 2,2).
Such two rotational temperatures were also observed in the M17SW 
molecular cloud (\markcite{Gus88}G\"{u}sten \& Fiebig 1988).

\subsection{Ammonia Abundance Enhancement}

The beam-averaged NH$_3$ column density in the quiescent gas at 
position A is estimated to be $N$(NH$_3$) = 2 $\times$ 10$^{14}$ cm$^{-2}$.
When we employed the H$_2$ column density of the ambient gas 
derived from the $^{13}$CO and C$^{18}$O observations ($N$(H$_2$) = 3$\times$10$^{21}$ cm$^{-1}$), 
the NH$_3$ abundance for the quiescent component is estimated to be 
$X$(NH$_3$) = 7 $\times$10$^{-8}$, 
which is comparable to those measured in nearby dense cores 
($X$(NH$_3$) = (3--10) $\times$10$^{-8}$; \markcite{Ben83}Benson \& Myers 1983).
The $N$(NH$_3$) in the blueshifted gas averaged over the beam at 
positions B and C are calculated to be 
1$\times$10$^{14}$ cm$^{-2}$ and 7$\times$10$^{13}$ cm$^{-2}$, respectively.
The H$_2$ column densities of the blueshifted gas were estimated from 
the CO ($J$=1--0) data 
obtained at the NRO 45m telescope (beam size was 16$''$) by convolving 
the data with the 72$''$ gaussian beam and assuming the optically thin CO emission, 
$T_{\rm ex}$ = 130 K, and H$_2$/CO abundance ratio of 10$^4$.
They are estimated to be 7$\times$10$^{21}$ cm$^{-2}$ toward position B 
and  2$\times$10$^{21}$ cm$^{-2}$ toward position C.
By using these H$_2$ column densities, 
we obtained $X$(NH$_3$) =  1$\times$10$^{-7}$ at position B 
and $X$(NH$_3$) = 3$\times$10$^{-7}$ at position C, which are a factor 
of 2--5 higher than that of the quiescent component.
The NH$_3$ column densities derived by 
\markcite{Taf95}Tafalla \& Bachiller (1995) from their high-resolution VLA data are 
$\sim$5$\times$10$^{14}$ cm$^{-2}$ toward both positions B and C.
If we compare these values with the $N$(H$_2$) in the 16$''$ beam 
derived from the CO ($J$=1--0) data by assuming an excitation 
temperature of 130 K, 
which are 2$\times$10$^{21}$ cm$^{-2}$ for position B and 
8$\times$10$^{20}$ cm$^{-2}$ for position C, 
we obtained $X$(NH$_3$) = 3$\times$10$^{-7}$ and 
$X$(NH$_3$) = 7$\times$10$^{-7}$ for positions B and C, respectively.
The H$_2$ column densities used here may be somewhat underestimated 
because the sizes of the NH$_3$ enhanced regions ($<$ 10$''$) 
in the map of \markcite{Taf95}Tafalla \& Bachiller (1995) are smaller 
than the beam size of the CO data.
When we take this into account, the ammonia abundances obtained from 
the higher resolution data are consistent with those from the 72$''$ resolution data.
Therefore, we conclude that the NH$_3$ abundance in the shocked regions 
is enhanced by a  factor of $\sim$5.

\subsection{Ortho-to-para Ratio and Its Implication for the Contribution of Desorbed Ammonia}

The derived ortho-to-para ratio in the blueshifted gas 
which is larger than the statistical value of 1.0
suggests that significant amount of NH$_3$ observed in the 
blueshifted gas has been evaporated from dust grains.
The evaporation of ammonia is supported by the fact that 
the ammonia abundance
is enhanced in the blueshifted gas.
If we assume that all of the NH$_3$ molecules observed in the 
blueshifted gas were from dust grains, 
the observed ratio of 1.3--1.7 suggest that the population of the 
rotational levels of NH$_3$ at the time of ejection from the 
grain surface is represented by a Boltzmann distribution 
with the temperature of 18--25 K.  
This formation temperature 
would be related to the excess energy distributed to the 
rotational degree of freedom in desorption processes, and are 
not necessarily equal to the dust temperature or the gas 
kinetic temperature 
(\markcite{Tak87}Takayanagi, Sakimoto, \& Onda 1987).
It is most likely that the NH$_3$ molecules retained on grains have been  
provided with 
sufficient energy to desorb from the grain surfaces by the passage of shocks 
(e.g. \markcite{Nej90,Flo95}Nejad et al. 1990; Flower et al. 1995).
The rotational temperature of 130--140 K suggests that 
the shock heating may be responsible for the desorption of the NH$_3$ 
molecules.
\markcite{San93}Sandford \& Allamandola (1993) 
revealed that the sublimation of ammonia drastically increases
as a function of temperature:
the residence time of NH$_3$ on a dust grain which is 10$^{13}$ yr at 
40 K becomes only 10$^{-7}$ yr at 100 K.

Recently, the infrared spectroscopic observations with Infrared Space 
Observatory (ISO) and ground-based telescopes reported 
that the abundance of the solid NH$_3$ (relative to H$_2$O) in icy grain 
mantle is no more than a few percent 
(e.g. \markcite{Whi96a,Whi96b}Whittet et al. 1996a, b).
These results imply that the NH$_3$ molecule desorbed from grains is 
less important than previously expected 
(e.g. \markcite{dHe85,Bro90,Nej90}d'Hendecourt et al. 1985; 
Brown \& Charnley 1990; Nejad et al. 1990).
However, the ortho-to-para ratio measured in the L1157 outflow 
indicates that significant amount of NH$_3$ arises from dust grains 
and that the gas-grain chemistry plays an important role in determining 
the NH$_3$ abundance.

\acknowledgments

We would like to thank the staff of NRO for the operation of our 
observations and their support in data reduction. 
We also thank Drs. S. Takano, K. Kawaguchi, Y. Hirahara, Y. Taniguchi,
and J. Tkahashi for helpful comments.
N.H. acknowledges support from a Grant-in-Aid from the Ministry of 
Education, Science, Sport, and Culture of Japan, No. 09640315.

\clearpage

\figcaption[fig1.eps]{A map of CO ($J$=1--0) outflow taken from Umemoto et al. (1992) with the observed positions marked.
Three positions marked by filled circles were observed with six transitions of NH$_3$ from (1, 1) to (6, 6) and those marked by open circles were observed with the (1, 1) to (3, 3) transitions.
Dashed contours indicate the distribution of the blueshifted emission (-5 to 1 km s$^{-1}$) and solid contours show the redshifted emission (4.5 to 10 km s$^{-1}$). \label{fig1}}

\figcaption[fig2.eps]{NH$_3$ spectra obtained at positions A (0$''$, 0$''$), B (20$''$, -60$''$), and C (45$''$, -105$''$).
The offsets are in arcseconds from the position of IRAS 20386+6751 [$\alpha$(1950) = 20$^{\rm h}$ 38$^{\rm m}$39.$^{\rm s}$6, $\delta$(1950) = 67$^{\circ}$51$'$33$''$]. \label{fig2}}

\figcaption[fig3.eps]{Rotation diagrams for the NH$_3$ transitions measured toward positions B (20$''$, -60$''$), and C (45$''$, -105$''$).
$N_{\rm u}$, $g_{\rm u}$, and $E_{\rm u}$ are the column density, the statistical weight, and the energy for the upper levels of the transitions.
Filled circles denote the para species and filled triangles are the ortho species.
Vertical error bars give 1$\sigma$ errors.
The solid lines between the (1, 1) and (2, 2) data points and the dashed lines between the (3, 3) and (6, 6) data points give the rotational temperatures derived from the combinations of the (1, 1) and (2, 2) data and of the (3, 3) and (6, 6) data.
If we assume an ortho-to-para ratio to be 1.7 for position B and 1.3 for position C (the column densities of the ortho species are denoted by the open triangles), the data points of four transitions from (3, 3) to (6, 6) are fitted by the solid straight lines. 
 \label{fig3}}

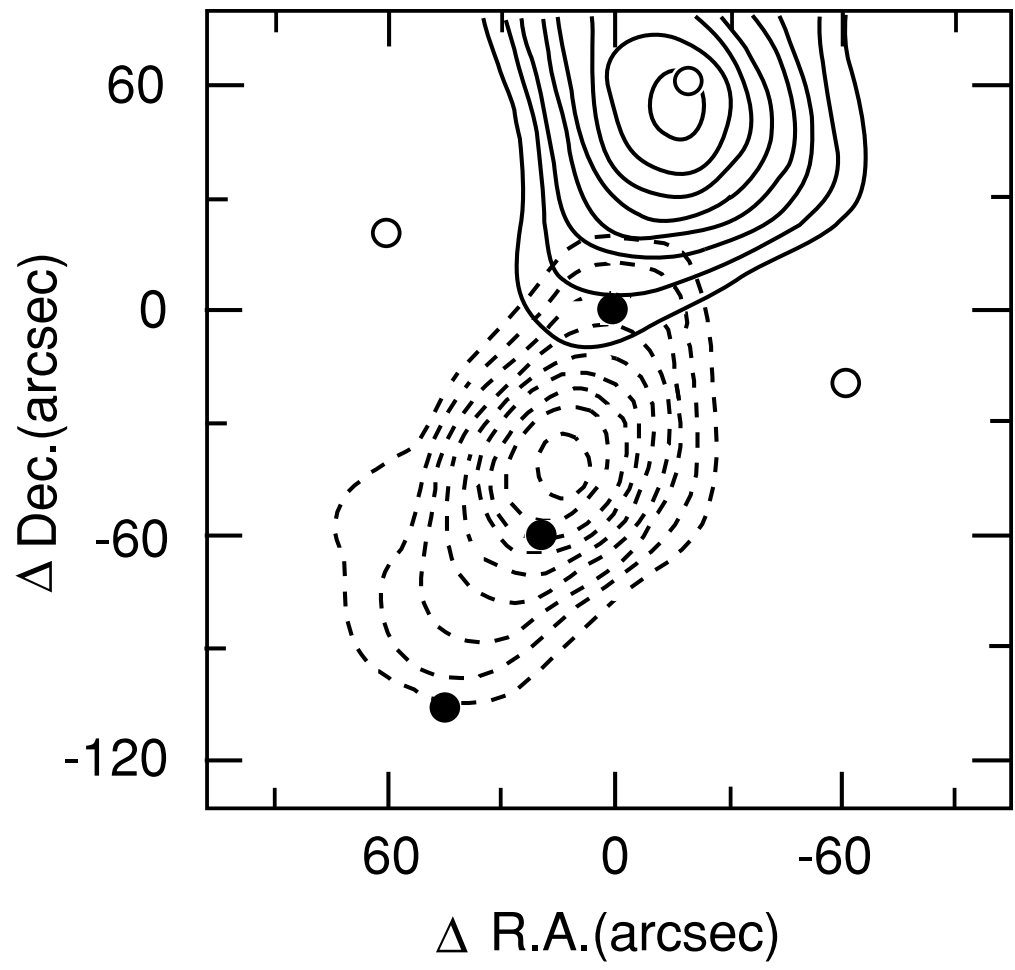
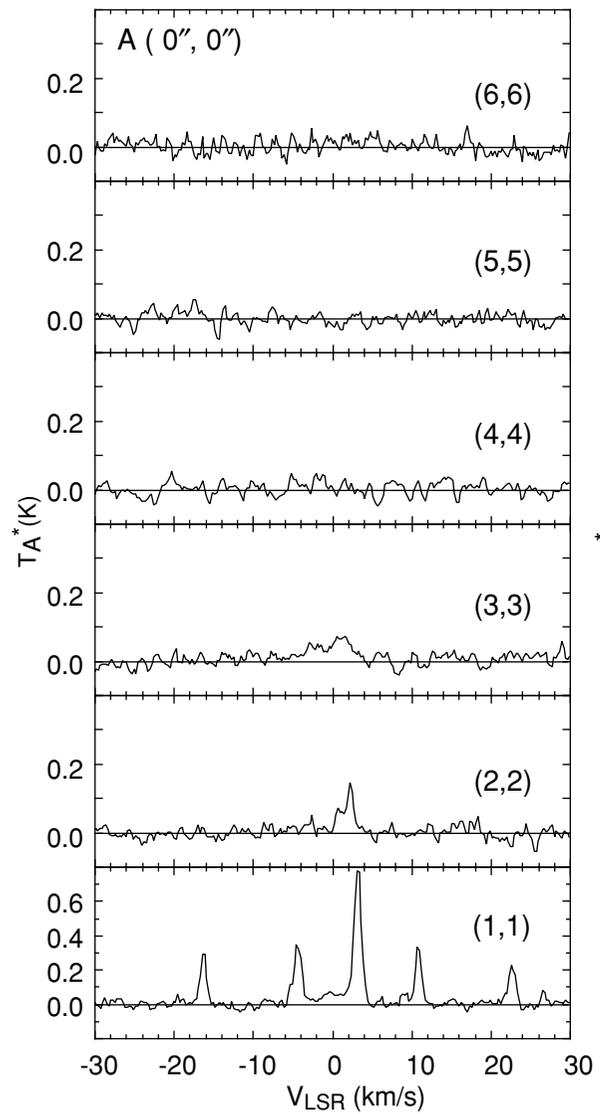
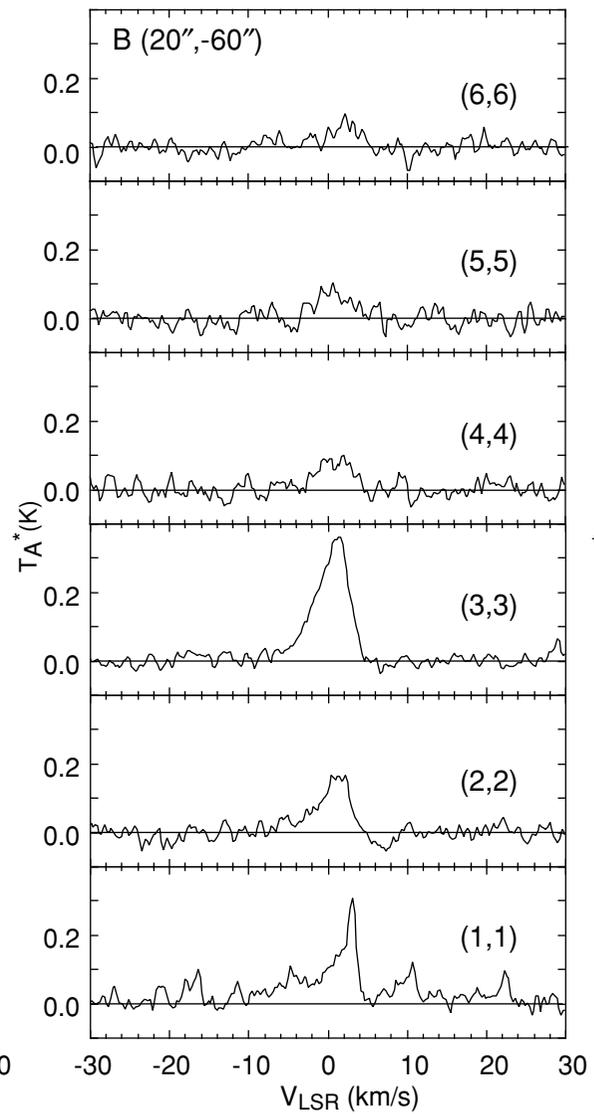
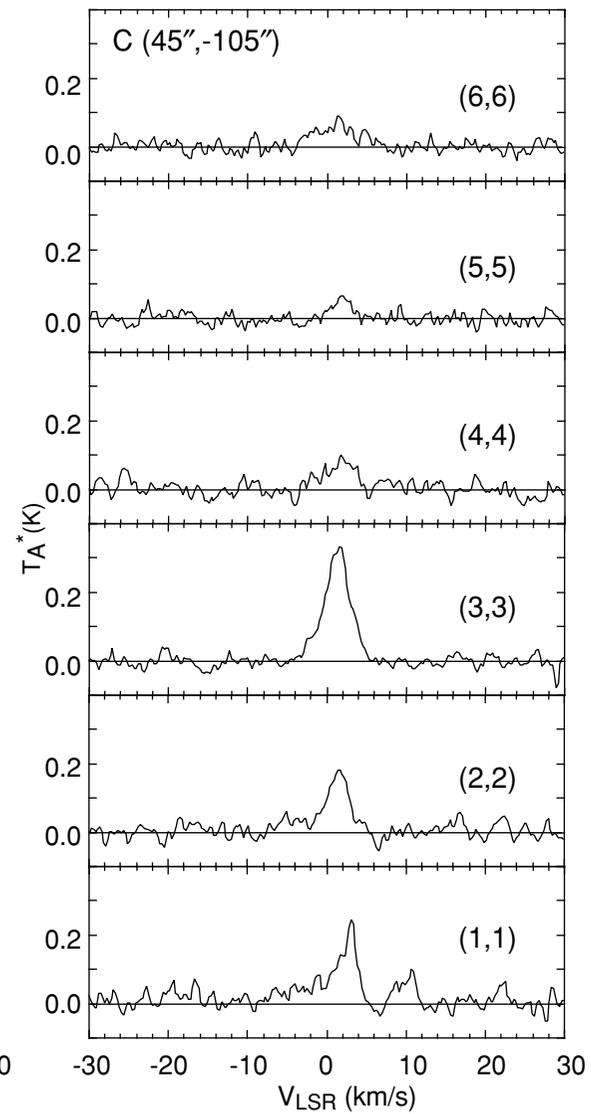
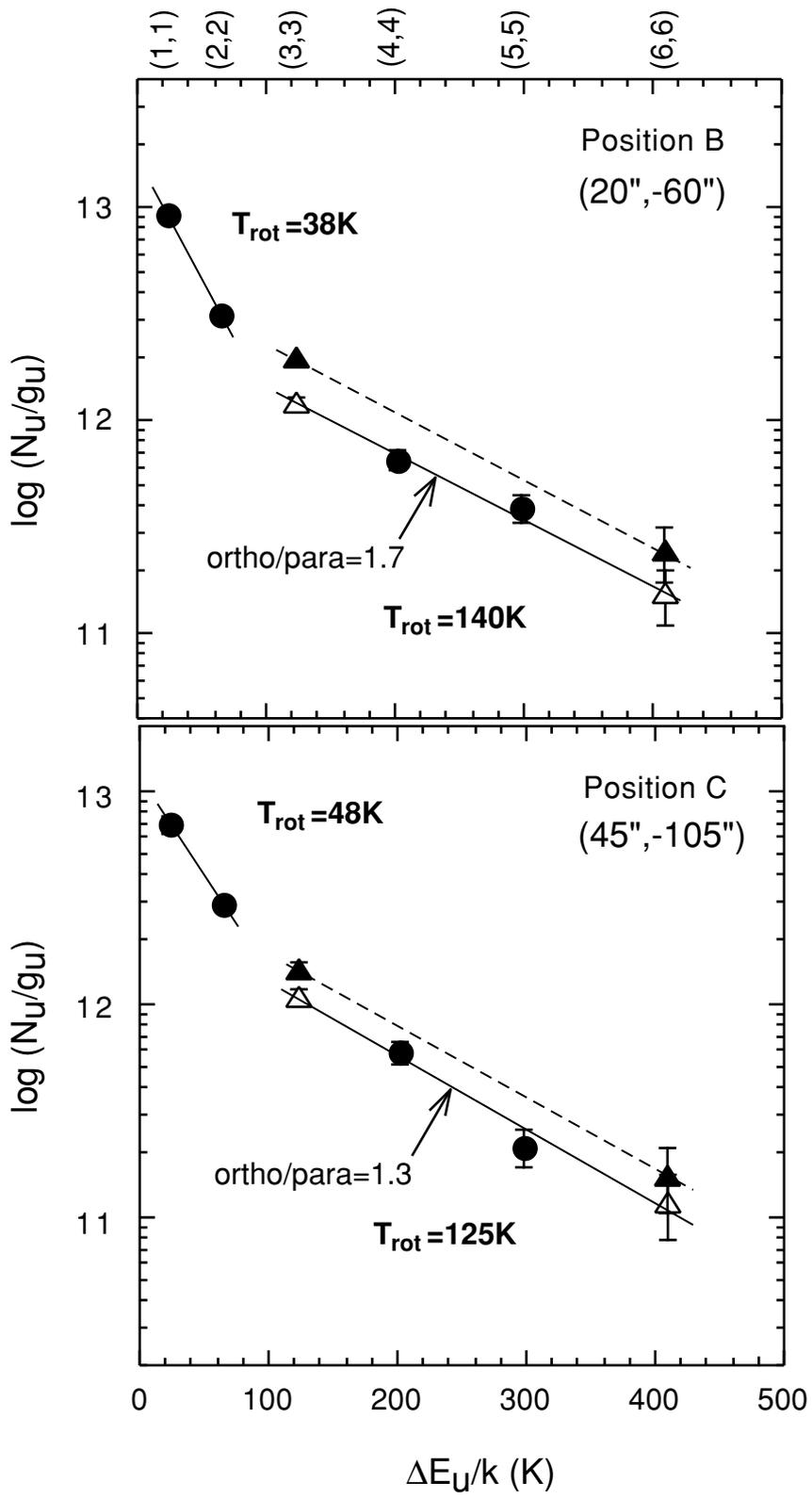
\end{document}